\documentstyle[11pt,moriond,epsfig]{article}

\def\be{\begin{equation}}
\def\ee{\end{equation}}
\def\bea{\begin{eqnarray}}
\def\eea{\end{eqnarray}}

\begin{document}
\vspace*{4cm}
\title{Two photon background for Higgs boson searches at the LHC}

\author{ T. Binoth }

\address{Laboratoire d'Annecy-le-Vieux de Physique Th\'eorique LAPTH}

\maketitle\abstracts{
The search for an intermediate mass Higgs boson at the LHC  
needs the quantitative understanding of the two-photon background. 
A calculation of two-photon production in hadronic collisions
at full next-to-leading order is described. It includes 
photons originating from the hadronization of QCD partons which
play an important role at the LHC. A prediction 
for the invariant mass distribution for photon pairs at 
the LHC is presented and finally
the residual scale dependencies are discussed.}

\section{Introduction}
The understanding of electroweak symmetry breaking is a major
motivation for present and future collider physics. Inside
the Standard Model (SM) the LHC is expected to detect the predicted 
Higgs boson, if its mass is below say 800 GeV. On one 
hand the direct search at the LEP experiments place a stringent lower
bound whereas on the other hand the precision measurements 
indicate an upper bound for the mass of the
Higgs boson \cite{muijs}:
\be
107.7 \,\mbox{GeV} < M_{Higgs} < 188 \, \mbox{GeV}  \qquad \mbox{(95\% c.l.)}      
\ee 
Also for the light Higgs boson of the minimal supersymmetric
standard model a lower bound of 88 GeV exists, where on the contrary 
the upper bound of around 130 GeV is dictated by the 
structure of the interaction terms. The quartic couplings are
related to the gauge couplings here, not a free parameter
as in the SM. The most promising signal for such neutral Higgs
bosons in the mass range of $80 \,\mbox{GeV} < M_{Higgs} < 140 \,\mbox{GeV}$
comes from the gluon fusion process with a subsequent decay
of the Higgs boson into a photon pair. Though the
branching fraction is small, 
$B(H\rightarrow \gamma\gamma) \sim 2 \times 10^{-3}$ in the SM,
one expects a sharp peak in the invariant mass distribution
of the photon pair, because of the narrow Higgs width of a few MeV. 
Though the signal is very sharp one is confronted
with a huge background. To get a quantitative understanding
of the signal to background ratio one has to know the background
as best as one can.

The background can be split into three contributions:
\begin{description}
\item[Direct:] Both photons originating from the hard
partonic interaction.
\item[Fragmentation:] At least one of the photons is created
in the hadronization of a QCD parton.
\item[Meson decay:] The photons are decay products of mesons
like $\pi^0$, $\eta$, and so on.
\end{description}     
Whereas the latter is essentially reducible the former are
commonly called irreducible backgrounds though one can
reduce the fragmentation contribution substantially by
imposing isolation criteria, i.e. one restricts the
hadronic energy inside a cone (defined in rapidity
and azimuthal angle space) around the photon momentum 
to be less than some value.

In the following the calculation of this background
at full next-to-leading order is shortly sketched. 
The inclusion of the NLO fragmentation contributions
is a step beyond the calculations of~\cite{abfs,owens,yuan}.
A prediction of the background $M_{\gamma\gamma}$
distribution at LHC is presented and its 
stability under variation of the different scales
entering the computation is discussed.    

\section{Photon pair production in hadronic collisions at NLO}

The Born term for photon pair production, $q\bar q\rightarrow \gamma\gamma$,
is of order $\alpha^2$. The 
radiative corrections consist out of all virtual 
and real emission corrections which lead to an order $\alpha^2\alpha_s$
correction.
One observes  that in subprocesses like $qg\rightarrow \gamma\gamma q$
a final state singularity is present if the photon is collinear
to the quark. To absorb these singularities into
a photon fragmentation function $D_{\gamma/q}$ one has subsequently
also to take into account other partonic reactions
like e.g. $qg\rightarrow \gamma q$. As $D_{\gamma/q}$
behaves like $\alpha/\alpha_s$ in the investigated kinematic regime,
the power counting of the couplings is the same as the one of the Born term.
As both photons can be collinear to external partons 
also both photons should be allowed to be  
created in hadronization. Two--fragmentation processes
are needed for consistency together with the respective
higher order corrections, see \cite{bgpw} for more details.
Finally there is another sizable contribution, $gg\rightarrow \gamma\gamma$,
where the gluons/photons are attached to a fermion loop.
Formally a higher order correction, this contribution is 
not negligible due to the large gluon flux at the LHC.\\
The diagrammatic calculation depends on three unphysical scales:
\begin{description}
\item[$\mu$:] Renormalization scale due to ultraviolet divergence,
\item[$\mu_{FACT}$:] Factorization scale due to QCD initial
state collinear singularities,
\item[$\mu_{FRAG}$:] Fragmentation scale due to final state 
collinear singularities.
\end{description}
The relation between the hadronic and the partonic cross sections
for a fragmentation process is done by folding the
partonic cross sections with the $\mu_{FACT}$-dependent parton distribution 
functions and the $\mu_{FRAG}$-dependent photon fragmentation functions 
$D_{\gamma / j \in \{ q,\bar q, g\}}$.
   
\section{The $M_{\gamma\gamma}$ distribution at the LHC}
\begin{figure}[t]
%\vspace{0.4cm}
\begin{center}
    \epsfxsize = 10.cm
    \epsffile{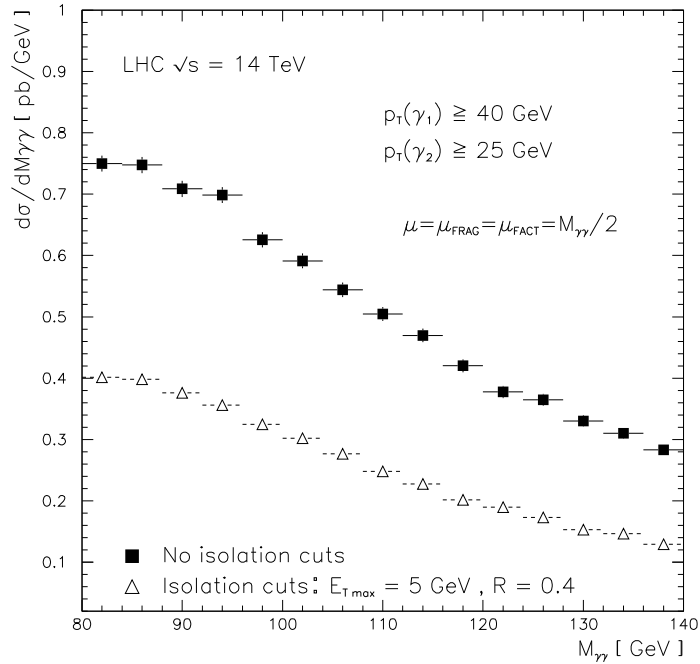}
\end{center}    
\caption{\label{Fig:1} The $M_{\gamma\gamma}$ distribution at the LHC
in the window relevant for Higgs searches using standard cuts. 
Applying no photon isolation criterion leads to the upper curve 
whereas the lower curve shows the effect of isolation. Whereas in
the upper curve the one-fragmentation contribution is dominant, in
the lower curve mainly the direct component is present.}
\end{figure}

\begin{figure}[t]
%\vspace{0.4cm}
\begin{center}
    \epsfxsize = 10.cm
    \epsffile{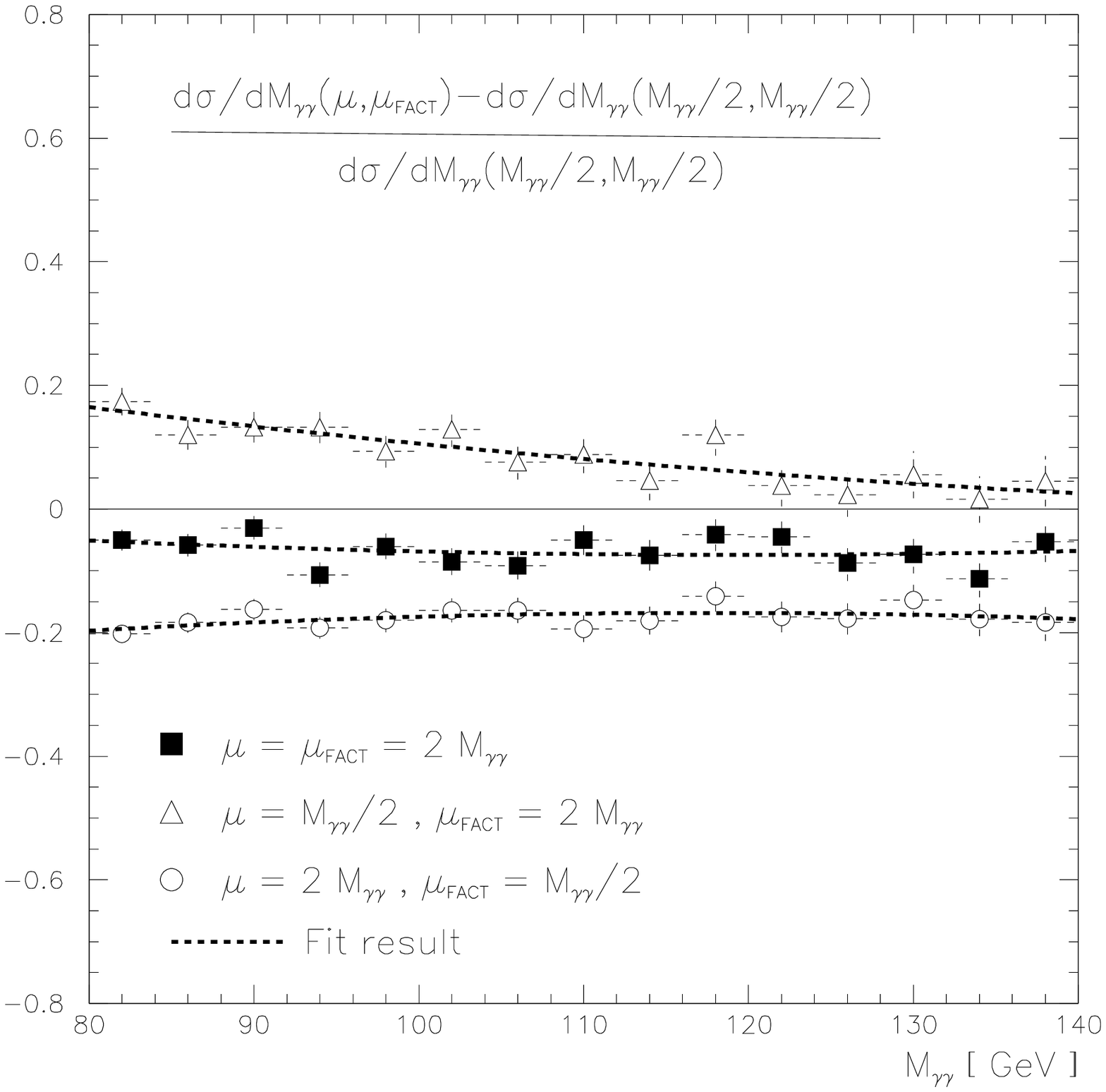}
\end{center}    
\caption{\label{Fig:2} The behaviour of the lower 
curve of Fig.~\ref{Fig:1} under scale variations. 
The relative difference is plotted for diagonal
and anti-diagonal variations of $\mu$ and $\mu_{FACT}$. 
The fragmentation scale is chosen as $\mu_{FRAG}=\mu_{FACT}$.}
\end{figure}

The comparison of our calculation with recent Tevatron data
is showing a good agreement~\cite{bgpw}. This is especially true for
infrared save quantities as the
$M_{\gamma\gamma}$ distribution~\footnote{Infrared sensitive 
observables are showing also good agreement for the
tails of the distributions. In the infrared sensitive region
resummation of large logarithms is needed \cite{yuan}.}.
In Fig.~\ref{Fig:1} we plot the invariant mass distribution
of a photon pair at the LHC with and without applying isolation cuts.
As an example we demand the transversal hadronic energy
to be less than $E_{T max}=5$ GeV in a cone $R=0.4$ around the photon 
defined in rapidity and azimuthal angle space through 
$R=\sqrt{(y-y_{\gamma})^2 + (\phi-\phi_\gamma)^2}$. 
Apart from these, standard cuts were taken as
indicated in the figure. The scale choice is
$\mu=\mu_{FACT}=\mu_{FRAG}=M_{\gamma\gamma}/2$.
For all figures the MRST2 set of parton distribution functions~\cite{mrs99} and 
the BFG photon fragmentation functions~\cite{phofrag} were used.
To get an idea of the residual scheme dependence 
of our result for the lower curve  in Fig.~\ref{Fig:1}
(with isolation), 
we varied the renormalization and
factorization scales and compared to that curve by plotting 
$[d\sigma/dM_{\gamma\gamma}(\mu,\mu_{FACT}) -
  d\sigma/dM_{\gamma\gamma}(M_{\gamma\gamma}/2,M_{\gamma\gamma}/2)  ]
 /[d\sigma/dM_{\gamma\gamma}(M_{\gamma\gamma}/2,M_{\gamma\gamma}/2)]$ 
in Fig.~\ref{Fig:2}.
Due to isolation the fragmentation scale dependence
is marginal~\footnote{This is because we consider  
the fragmentation contributions in the calculation. 
Thus the result is of beyond leading logarithmic accuracy.}
and is kept equal to $\mu_{FACT}$ for simplicity.
In Fig.~\ref{Fig:2} one observes that there is an 
accidental stability if $\mu,\mu_{FACT}$
are changed in the same directions from 
$\mu=\mu_{FACT}=M_{\gamma\gamma}/2$ to 
$\mu=\mu_{FACT}=2 M_{\gamma\gamma}$. This is due to the fact that
at LHC for the given parton energies the effect of larger (smaller)
$\mu$ -- means smaller (larger) NLO corrections --
is compensated by the increased (decreased) gluon flux if $\mu_{FRAG}$
is also increased (decreased). 
By varying in an anti-diagonal way one gets a more conservative
estimation of the uncertainty. One observes variations
between 10 and 20 per cent.
\section{Conclusion}
A calculation for the photon pair production at hadron
colliders at full NLO  exists and is implemented into
a partonic event generator, called DIPHOX \cite{bgpw}.
Tevatron data are well described by the code and
predictions for the LHC show a residual scale dependence 
of the order of 10 to 20 per cent~\footnote{To get a further stabilization 
at least the dominant processes would have to be calculated
one order higher. E.g. the corrections to the box contribution
have to be included but also the corrections 
to  the gluon induced subprocesses.}.  
For infrared sensitive observables, like e.g. $P_T$ distributions, 
resummation has to be included in the calculation. 
In contrary the $M_{\gamma\gamma}$ distribution is now known 
at NLO not only in the direct but also in the fragmentation part
which allows more reliable quantitative studies of effects due 
to isolation criteria.

\section*{Acknowledgments}
I would like to thank my collaborators J.~Ph.~Guillet, E.~Pilon and
M.~Werlen for giving me the opportunity to present
our results at the Moriond conference.\\ 
This work was supported by the EU Fourth Training Programme  
''Training and Mobility of Researchers'', Network ''Quantum Chromodynamics
and the Deep Structure of Elementary Particles'',
contract FMRX--CT98--0194 (DG 12 - MIHT).

\end{document}